\documentclass[structabstract]{aa}  
\usepackage{graphicx}
\usepackage{natbib}
\usepackage{txfonts}
\begin{document}
   \title{Accurate photometry with adaptive optics in the presence of
     anisoplanatic effects with a sparsely sampled PSF}

   \subtitle{The Galactic center as an example of a challenging target for
   accurate AO photometry}

   \author{Rainer Sch{\"o}del
          \inst{1}
          }

   \institute{Instituto de Astrof\'isica de Andaluc\'ia (CSIC), C/
     Camino Bajo de Hu\'etor 50, 18008 Granada, Spain\\
              \email{rainer@iaa.es}
}

  \authorrunning{R. Sch{\"o}del}
  \titlerunning{AO photometry with sparsely sampled PSF}
   \date{;;}

 
  \abstract
  {Anisoplanatic effects can cause significant systematic photometric
    uncertainty in the analysis of dense stellar fields observed with
    adaptive optics. Program packages have been developed for a
    spatially variable PSF, but they require that a sufficient number of
    bright, isolated stars in the image are present to adequately sample
    the PSF.}
   {Imaging the Galactic center is particularly
     challenging. We present two ways of dealing with
     spatially variable PSFs when only one or very few
     suitable PSF reference stars are present in the field.}
   {Local PSF fitting with the {\emph{StarFinder}} algorithm is
     applied to the data. Satisfying results can be found in two ways:
     (a) creating local PSFs by merging locally extracted PSF cores
     with the PSF wings estimated from the brightest star in the field;
     (b) Wiener deconvolution of the image with the PSF estimated from
     the brightest star in the field and subsequent estimation of
     local PSFs on the deconvolved image. The methodology is tested on
     real, and on artificial images.}
{The method involving Wiener deconvolution of the image prior to local
  PSF extraction and fitting gives excellent results.  It limits
  systematic effects to $\sim2-5\%$ in point source photometry and
  $\sim10\%$ in diffuse emission on fields-of-view as large as
  $28"\times28"$ and observed through the $H$-band filter.  Particular
  attention is given to how deconvolution changes the noise
  properties of the image. It is shown that mean positions and fluxes
  of the stars are conserved by the deconvolution. However, the
  estimated uncertainties of the PSF fitting algorithm are too
  small if the presence of covariances is ignored in the PSF fitting
  as has been done here. An appropriate scaling factor can,
  however, be determined from simulated images or by comparing
  measurements on independent data sets.  }
   {We present ways of obtaining reliable photometry and astrometry from
  images with a spatially variable, but poorly sampled PSF, where
  standard techniques may not work. }


   \maketitle

%

\section{Introduction}

Aperture photometry is clearly the most straightforward and -- when
examining isolated sources -- most accurate means of determining photometry.
Nevertheless, it runs into serious difficulty in crowded
fields. Frequently, adaptive optics (AO) is applied to near-infrared
observations of crowded stellar fields. This reduces source
confusion considerably, but the limited Strehl ratio leads to extended
wings of the point spread function (PSF).  As a result, the complete
PSF usually has a diameter of the order of the seeing disk, even if a
large fraction of the flux is concentrated inside a circular region with a
radius of at most a few times the diffraction limit. Therefore, the
light from stars in dense fields will overlap even when using AO
techniques.  For this reason, crowded stellar fields are usually
analyzed with PSF fitting program packages, such as DAOPHOT
\citep{Stetson:1987nx} or {\it StarFinder}
\citep{Diolaiti:2000rz}. The latter program package was written
explicitly for use with AO data and uses an empirical PSF that is
extracted directly from the data.

If the PSF is constant over the field, the astrometric and photometric
accuracy of PSF fitting algorithms is limited  only by the
signal-to-noise ratio of the imaging data and the accuracy of one's
knowledge of the PSF. However, the PSF is never constant across the
field-of-view (FOV), but subject to changes caused by distortions and
aberrations in the optical path. A particularly strong effect is the
change of the PSF as a function of distance from the guide star in AO
observations. This effect, termed anisoplanacy, is due to 
different lines-of-sight probing different parts of the turbulent
atmosphere. Various methods have been developed for taking anisoplanatic
effects into account a posteriori. Some of these methods involve, e.g.,
source fitting with local PSF estimates on subfields smaller than the
isoplanatic angle \citep[e.g.,][]{Diolaiti:2000rz,Christou:2004gf},
space-variant deconvolution \citep[e.g.,][]{Diolaiti:2000rz},
analytical formulations of anisoplanatism combined with the guide star
PSF and information about the atmospheric turbulence profile
\citep{Britton:2006kx}, or semi-empirical PSF modeling based on
observations of dense stellar fields \citep{Steinbring:2002ul}. The
PSF fitting program package DAOPHOT deals with variable PSFs by
allowing for up to a quadratic variability in mathematical PSF models
and combining those with local look-up tables
\citep[e.g.,][]{Stetson:1992eu}.  Multiconjugate
\cite[MCAO,][]{Beckers:1988lq} or multiobject AO
\citep{Hammer:2004dq,Ellerbroek:2005rr} techniques will reduce the
effects of anisoplanatism a priori in future observations,
i.e., before the light reaches the detector. The multiconjugate AO
technique has already been successfully demonstrated on the ESO VLT
using the Multiconjugate Adaptive Optics Demonstrator
\citep[e.g.,][]{Bouy:2008qf,Gullieuszik:2008pd,Marchetti:2008bh}. Nevertheless,
while MCAO improves the Strehl ratio of sources  dramatically over a
wide FOV, significant and difficult to predict PSF variations across the
FOV remain. Dealing with spatially variable PSFs in
photometry is therefore important and will remain so
in the near future.

All approaches to PSF fitting with spatially variable PSFs need a
  sufficient number of bright, isolated PSF reference stars that are
  approximately homogeneously distributed across the FOV. In other
  words, the PSF and its spatial variation must be adequately sampled.
  In this paper, I present a data set that poses particular problems for
  this latter assumption. Two ways are proposed on how to overcome the
  difficulties. The most successful method consists  of a combination
of linear, Wiener-filter deconvolution and local PSF fitting. The
image is first Wiener-filter-deconvolved using a suitable PSF, ideally
the one of the guiding star, to reduce crowding. In a second step, the
local variation in the PSF and the ringing introduced by the Wiener
filter technique is taken care of by PSF fitting with locally
extracted PSFs.  The method is easy to implement. Deconvolution
  will lead to co-variances in the noise. It is shown that this does
  not lead to erroneous measurements for  Wiener deconvolution
  and  the imaging data used here. However, care must be taken to
  obtain accurate estimates of the uncertainties in the positions and
  fluxes.

  The algorithm was developed and tested on near-infrared observations
  of the crowded Galactic center field with NACO at the ESO VLT and
  should be applicable in general to AO observations of crowded
  fields. An important prerequisite however, is a sufficient density
  of point sources over the entire FOV to be able to estimate
  the PSF after deconvolution with sufficient and nearly constant
  quality across the field. Applied to the data analyzed in this work,
  systematic variations of point source photometry out to distances of
  $\sim30"$ from the guide star can be limited to
  $\lesssim\,2-5\%$. Additionally, the method presented does not only
  lead to accurate point source photometry, but also allows one to
  determine the diffuse emission due to unresolved stars with a
  $1\sigma$ accuracy of $\sim\,0.1$\,mag\,arcsec$^{-2}$ across the
  entire FOV.


\section{Data \label{sec:data}}

The observations used in this work were obtained with the
near-infrared camera and AO system NAOS/CONICA (short
NACO) at the ESO VLT unit telescope~4 \footnote{Based on observations
  made with ESO Telescopes at the La Silla or Paranal Observatories
  under programme ID 077.B-0014}. The $\rm mag_{Ks}\approx6.5-7.0$
supergiant IRS\,7  was used to close the loop of the AO,
using the instrument's unique NIR wavefront sensor. The sky
background was measured on a largely empty patch of sky, a dark cloud
about $400''$ north and $713''$ east of the target. Sky subtraction,
bad pixel correction, and flat fielding were applied to the individual
exposures. The NACO S27 camera, with a pixel scale of $0.027''/{\rm
  pix}$, was used for the $H$-band observations. The
field-of-view (FOV) of a single exposure is thus $28''\times28''$. The
observations were dithered by applying a rectangular dither pattern
with the center of the dithered exposures positioned approximately at
 $(8.0'',-2.6'')$, $(-6.1'',-2.7'')$, $(-6.1'',11.2'')$, and $(8.1'',11.3'')$
 east and north of Sgr\,A*. In the text we refer to
these four offsets as dither positions 1, 2, 3, and 4. The combined FOV of the
observations is about $40''\times40''$ and is offset to the north with
respect to Sgr\,A* because the guide star IRS\,7 is located about
$5.6''$ north of Sgr\,A*.  

\begin{figure}[!htb]
\includegraphics[width=\columnwidth,angle=0]{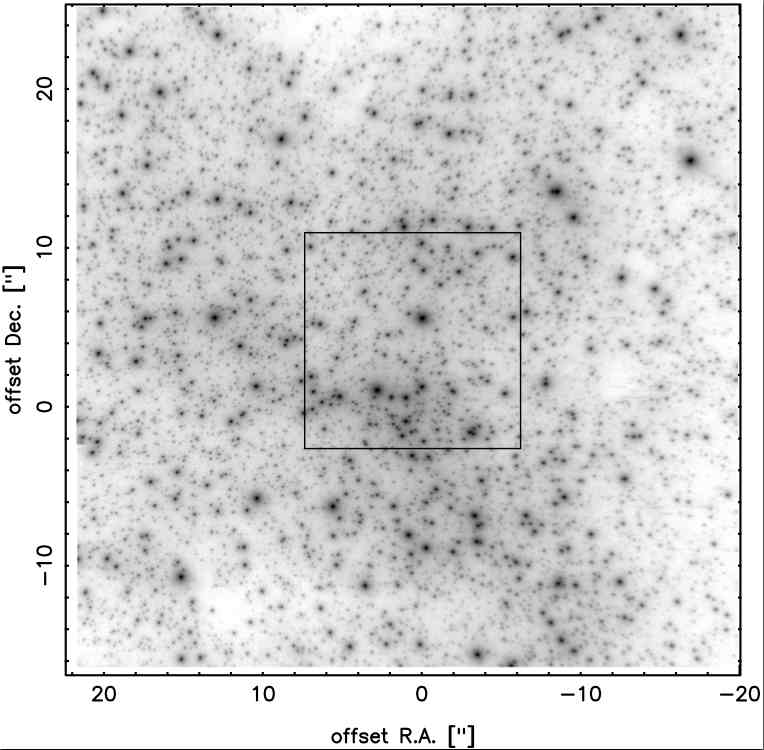}
\caption{\label{Fig:mosaic} Mosaic image of the $H$-band observations from 29
  April 2006. North is up and east is to the
  left. Offsets are given in arcseconds from Sgr\,A*. The black
  rectangle indicates the area of overlap between the 4 dither positions.} 
\end{figure}

Seeing ranged between $0.6"$ and $1.0"$. The Strehl ratio that was achieved
ranged between $\sim15\%$ near the guide star and $\sim8\%$ at $25"$
distance from the guide star. The Strehl ratio was estimated using the
{\it Strehl} algorithm of the ESO {\it eclipse} software package
\citep{Devillard:1997kx} on PSFs extracted at various positions in the
image. From the multiple measurements, we estimated the $1\,\sigma$
uncertainty of the measured Strehl ratio to be $\sim3\%$.  Sky
transparency variations were below $1\%$ during the observations.
Table\,\ref{Tab:Obs} summarizes the observations.  The detector
integration time (DIT) was set to be $2.0$\,s to avoid
saturation of the brightest stars. After 28\,DITs, the instrument
averaged the data to a single exposure (NDIT$=28$). In this way, 8
individual exposures were obtained per dither position. The exposures
of each respective dither position were aligned (to compensate for
small residual shifts) with the {\it jitter} algorithm of the ESO {\it
  eclipse} software package. We show the combined FOV of the $H$-band
observations in Fig.\,\ref{Fig:mosaic}. We note that the photometry and
astrometry were performed on the combined images of {\it each
  dither position} and {\it not on the combined mosaic} of all images
(as shown in Fig.\,\ref{Fig:mosaic}) to maintain a constant
signal-to-noise ratio over the entire images. The
$\sim13.5"\times13.5"$ overlap area between the four dither positions
is indicated by the rectangle in Fig.\,\ref{Fig:mosaic}.

\begin{table}
\caption{Details of the imaging observations used in this
  work.}
\label{Tab:Obs} 
\begin{tabular}{llllll}
\hline
\hline
Date & $\lambda_{\rm central}$ & $\Delta\lambda$ & N$^{\mathrm{a}}$ & NDIT$^{\mathrm{b}}$ & DIT$^{\mathrm{c}}$\\
 &  [$\mu$m]  &   [$\mu$m] &  & & [s] \\
\hline
29 April 2006 & 1.66 & 0.33 & 32 & 28 & 2  \\
\hline
\end{tabular}
\begin{list}{}{}
\item[$^{\mathrm{a}}$] Number of (dithered) exposures
\item[$^{\mathrm{b}}$] Number of integrations that were averaged on-line by the read-out
  electronics
\item[$^{\mathrm{c}}$] Detector integration time. The total integration time of each observation amounts to N$\times$NDIT$\times$DIT.
\end{list}
 \end{table}

\begin{figure*}[!htb]
\includegraphics[width=.95\textwidth]{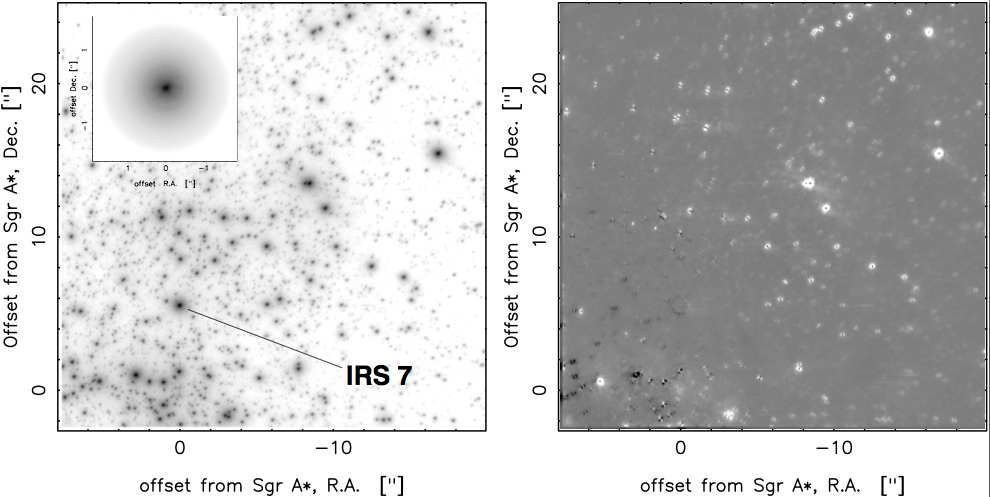}
\caption{\label{Fig:psfextract} Left: Image from dither position 3
  (see section\,\ref{sec:data}). The gray scale is logarithmic.  The
  position of IRS\,7, the guide star, and the PSF reference, is
  indicated. The inset shows the PSF extracted from IRS\,7 on a
  logarithmic gray scale. Right: Residuals (linear gray scale) after
  analyzing the image with the PSF extracted from IRS\,7. Here, the gray
  scale is linear and has been clipped for clearer illustration.}
\end{figure*}

Zero points for the NACO instrument for various combinations of
cameras, filters, and dichroics (that divide the light between
wavefront sensor and camera) are determined routinely within the ESO
instrument calibration plan. The zero point for the $H$-band
and for the corresponding setup (camera S27, dichroic N20C80) was
determined by observing a standard star during the same
night as the observations: $ZP_{H}=23.64\pm0.05$.

Anisoplanatic effects are stronger at shorter wavelengths, therefore
I chose $H$-band data (instead of $Ks$, the usual filter for observing
the Galactic center because of the strong extinction) to test the
photometric methodology presented in this work.


\section{PSF fitting photometry and astrometry \label{sec:data}}

Photometry and astrometry in crowded fields is usually performed with a PSF
fitting algorithm. Perhaps the most widely used software of this kind
is DAOPHOT \citep[e.g.,][]{Stetson:1987nx,Stetson:1992eu}, which is
also integrated in the {\it IRAF} software package. Another popular
package is SExtractor \citep{Bertin:1996oq}.  In this work, I use the
{\it StarFinder} algorithm \citep{Diolaiti:2000qo}, which was
specifically developed for images obtained by AO assisted
observations and is fairly popular in the AO community. It has been
shown to obtain comparable results to those of DAOPHOT \citep[in the
isoplanatic case, see][]{Diolaiti:2000qo,Diolaiti:2000rz}. I also
experimented with DAOPHOT (see section\,\ref{sec:daophot}).

In the {\it StarFinder} algorithm, an empirical point spread function
(PSF) is extracted by using one or several stars in the image. Cross
correlation on potential stars is performed. A correlation threshold
is set to accept/reject potential stars. Gaussian readout noise
and Poissonian photon noise are determined by the algorithm and taken
into account in the fitting process to determine formal
uncertainties. A smooth diffuse background emission is fit for the
image simultaneously to the point source photometry and
astrometry. The PSF extraction can be iteratively improved by using
the measured positions and fluxes of detected stars for removal of
secondary sources near the PSF reference stars.

There are many parameters that can be modified in the {\it StarFinder}
algorithm. The most important ones are the number of iterations and
the point source detection threshold applied at each iteration, the
size of the box for background estimation, and the correlation
threshold.  The parameter $thresh$ gives both the number of
iterations and the threshold in terms of standard deviations from the
noise, e.g., $thresh=[3.,3.]$ means two iterations with a $3\,\sigma$
threshold for each one. The parameter $back\_box$ is given in pixels
and designates a box size. {\it StarFinder} estimates the sky
background in boxes of size $back\_box\times back\_box$ and then
computes a smooth background by means of interpolation between the background
grid points. By default, {\it StarFinder} applies bilinear
interpolation. We  found that this tends to overestimate the
background close to bright point sources and therefore chose cubic
interpolation. The minimum required value for a correlation to exist between a
point source and the PSF is given by $min\_corr$. In this work, the
following values of these parameters were applied: $thresh=[5.]$
for a first detection of sources that are subsequently used to
iteratively improve the PSF;  $thresh=[5.,5.]$ for point source
extraction; $back\_box=60$ ($back\_box=30$ on deconvolved
  images); and $min\_corr=0.8$ ($min\_corr=0.9$ on deconvolved images).

For accurate photometry and astrometry, a number of effects
must be taken into account in addition to this basic
methodology. We identified the following points as being particularly
important:
\begin{itemize}
\item Accuracy of the empirically determined PSF.
\item Variation in the PSF across the field.
\item Influence  of deconvolution techniques on the results.
\end{itemize}
The last point is in part a consequence of the first two points and
the desire to improve the accuracy of the applied methodology. In the
following subsections different aspects of the PSF fitting method
are addressed in detail.

In this section, we address the points raised in the above
list. We use $H$-band imaging data in our analysis of
anisoplanatic effects because the variation in the PSF across the
field is normally stronger at shorter wavelengths. Some concern may be
raised because the ideal diffraction-limited PSF of the VLT in the
$H$-band is barely Nyquist sampled with the $0.027"$ per pixel camera
scale. However, the FWHM of the PSFs over the entire FOV is
$>3$\,pixels in all the images. This means that the PSF is sufficiently well
sampled and the fitting algorithm applied by {\it StarFinder} works
reliably, as we have also been able to confirm with simulated images
(see section\,\ref{sec:simulation}).


\subsection{PSF extraction}

The {\it StarFinder} algorithm uses of an empirical PSF, i.e., a
PSF directly extracted from the imaging data. This is recommendable in
AO observations because of the complexity of the PSF (e.g., a partial,
possibly broken, airy pattern superposed on a Gaussian seeing disk,
and speckles in the PSF wings) that cannot be easily described by 
mathematical models. The PSF in {\it StarFinder} is determined from
the median superposition of various stars or, alternatively, from a
single, bright, isolated, and unsaturated star.  Knowledge of the
empirical PSF is limited because of the limited S/N of the imaging
data and the presence of secondary sources in the wings of the PSFs of
the reference stars.  Radial and angular smoothing of the wings of the
empirically determined PSF is implemented to improve the S/N
in the faint wings.

The optimal PSF reference star is as bright as possible without being
saturated and completely isolated, i.e., with no secondary sources
close to the core or in the seeing wings.  These requirements are not
easy to fulfill in crowded fields. In the case of the Galactic center
(GC), the supergiant IRS\,7 has been used routinely as a PSF reference
star for GC observations with NACO since the year
2002. Interferometric observations with the ESO VLT Interferometer
(VLTI) confirm that this supergiant is unresolved at 2\,$\mu$m with an
aperture of 8\,m \citep{Pott:2008kx}. In the near-infrared, any star
within $0.5"$ of IRS\,7 is $>4$ magnitudes fainter (even
taking into account the variability of IRS\,7; see
\citealt{Blum:1996mz} and \citealt{Ott:1999ly}). The brightest star within $1"$ of
IRS\,7 is $>3$ magnitude fainter than IRS\,7. An iterative
approach is implemented in the {\it StarFinder} code that helps us to
effectively remove secondary sources in the PSF estimation process.
IRS\,7 is therefore well suited to providing a  PSF reference.

In the left panel of Fig.\,\ref{Fig:psfextract}, we show the image
corresponding to dither position 3 (see section\,\ref{sec:data}). The
inset shows the PSF extracted from IRS\,7. Besides the iterative
approach, the S/N in the faint wings was improved by applying
the HALO\_SMOOTH routine provided by {\it StarFinder} (angular width
set to $180$\,deg, radial width to 20\,pixels). As can be seen, the
PSF is clearly defined out to distances $>1"$ from the PSF core. The
right panel of Fig.\,\ref{Fig:psfextract} shows the residuals
\footnote{In this work, residuals are defined as $image - point\,
  sources - smooth\, diffuse\, emission$, such as given by
  StarFinder. This means that the residual image should ideally
  fluctuate around zero. Negative (positive) residuals around bright
  stars will have equivalent positive (negative) regions of diffuse
  background emission associated with them. } across
the image that result after fitting the point sources and the diffuse
emission in the image using the PSF extracted from IRS\,7. For clearer
illustration, the linear gray scale of the image of the residuals has
been clipped. As can be seen, the residuals related to point sources
vary systematically across the FOV. This demonstrates how using a
single, fixed PSF leads inevitably to systematic errors in the
photometry when the FOV is larger than the isoplanatic angle.

\begin{figure}[!htb]
\includegraphics[width=\columnwidth]{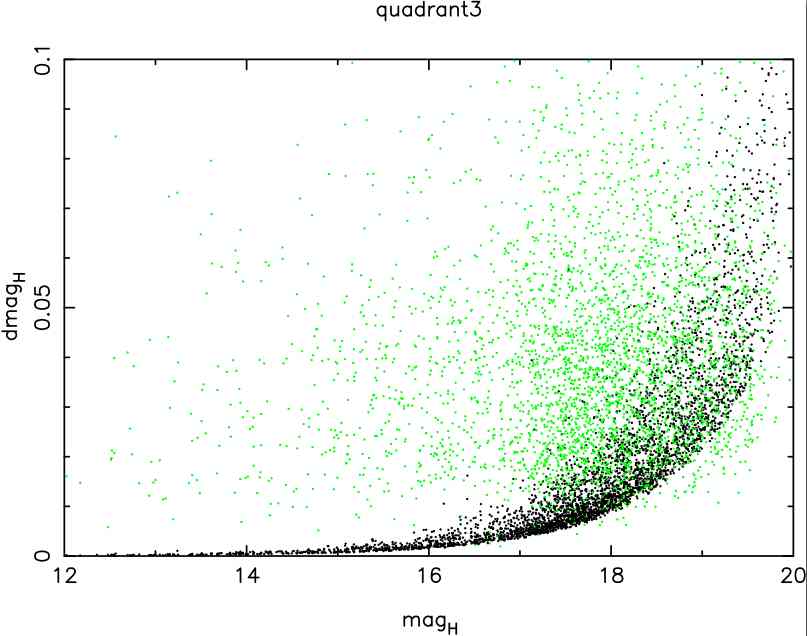}
\caption{\label{Fig:dphotloc} Uncertainty in PSF fitting photometry
  with locally extracted PSFs in the image from dither position
  3. Black dots are the formal photometric uncertainties computed by
  the {\it StarFinder} algorithm. Green dots are the photometric
  uncertainties due to the uncertainty in the PSF estimate. }
\end{figure}

\begin{figure*}[!htb]
\includegraphics[width=\textwidth]{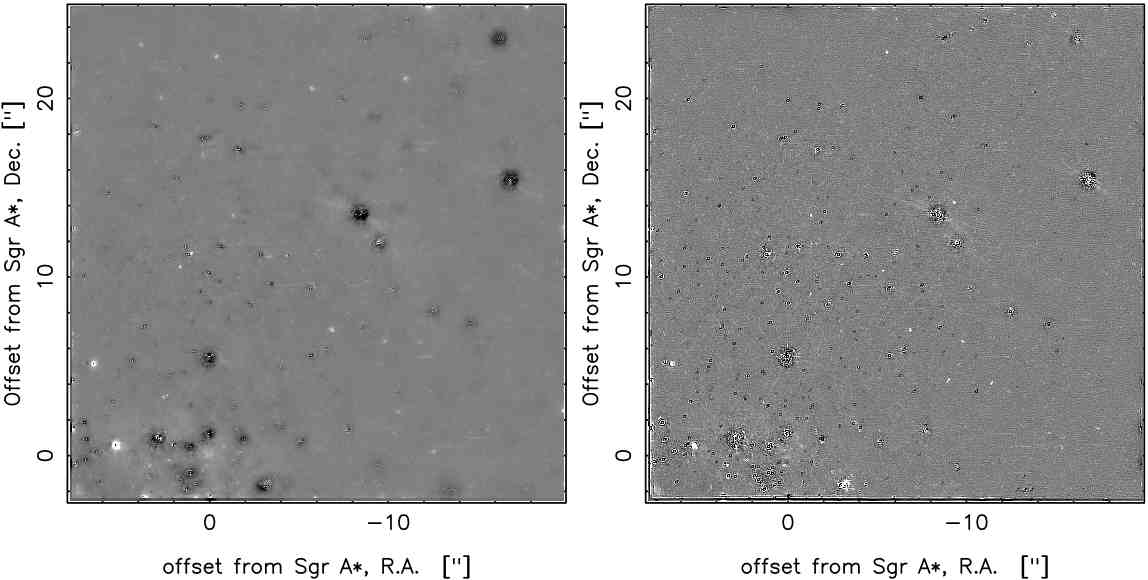}
\caption{\label{Fig:residlinloc} Left: Residuals (linear gray scale)
  after analyzing the image from dither position 3 with locally
  extracted PSFs.  Right:  Residuals after LW deconvolution of the
  image from dither position 3 with the guide star PSF and local PSF
  fitting. The gray scale in both panels is identical to the one applied to the right
  panel of Fig.\,\ref{Fig:psfextract}.}
\end{figure*}


\subsection{Local extraction of the PSF  \label{sec:spatial}}

The FOV of the NACO S27 camera ($28"\times28"$) is larger than the
isoplanatic angle, which is of the order $\sim10"$ in the $H$-band,
but depending strongly on the momentary atmospheric conditions. This
leads to a drop in the Strehl ratio and a change of the shape of the
PSF with distance from the guide star. Usually the PSF appears
elongated, the long axis pointing toward the guide star. By using a
single PSF in the analysis of extensive AO observations with a large
FOV, systematic errors in both photometry and astrometry over the
entire image can therefore be produced (see right panel of
Fig.\,\ref{Fig:psfextract}). At large distances from the guide star,
anisoplanatic effects can even cause the detection of spurious
sources because PSF fitting algorithms such as {\it StarFinder } may
try to separate elongated sources into two or more stars
\citep[see][]{Schodel:2007tw}.

\begin{figure*}[!htb]
\includegraphics[width=.95\textwidth]{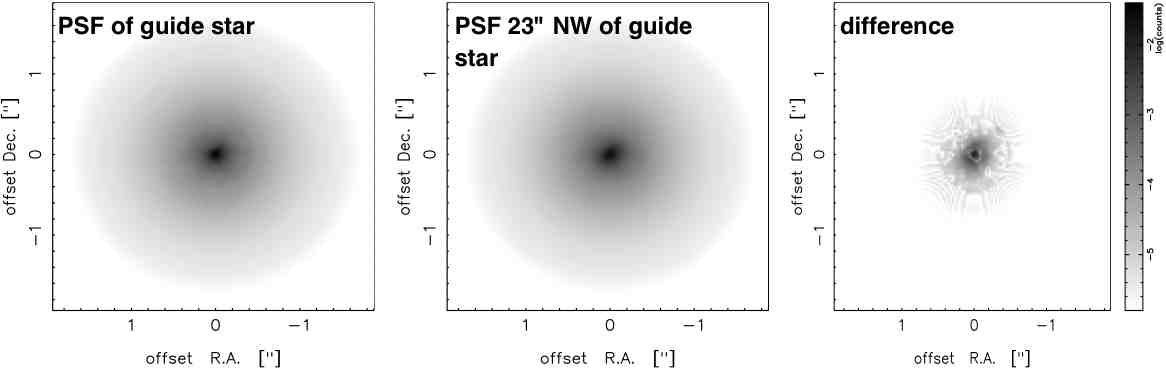}
\caption{\label{Fig:PSFs} Left: PSF at the position of the guide
  star. Middle: PSF $\sim23"$ NW of the guide star. Right: Difference
  between PSFs in left and middle panels. The logarithmic gray scale
  is identical for all images and is indicated in the bar that
  accompanies the right panel. The circular feature that can be seen
  in the left and right panels is the radius beyond which the guide
  star PSF was radially and azimuthally smoothed.}
\end{figure*}

\begin{figure*}[!htb]
\includegraphics[width=.95\textwidth]{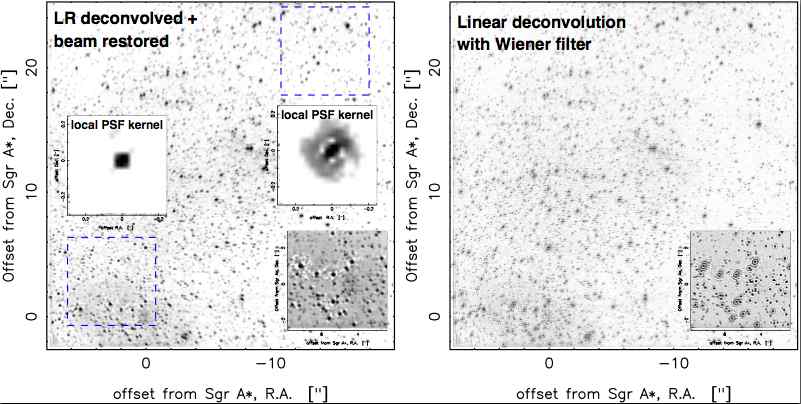}
\caption{\label{Fig:deconv} Left: $H$-band image from dither position
  3 after Lucy-Richard deconvolution and beam restoration with a beam
  of 3 pixel FWHM. Right: Linear (Wiener filter) deconvolution of the
  same image.  The insets in the left figure show the local PSF
  kernels extracted from $6.9"\times6.9"$ regions in two different
  fields of the LR deconvolved image (dashed boxes). The insets at the
  lower right corner of the two images show details near Sgr\,A*. Note
  that the typical ringing can be seen around the point sources in the
  Wiener-deconvolved image.}
\end{figure*}

An obvious way to take anisoplanatic effects into account is by using
{\it locally} extracted PSFs. A prerequisite for this technique is a
sufficiently high density of point sources across the FOV. This
condition is clearly fulfilled in the GC observations analyzed here.
However, while there are many point sources all over the field, there
exist large patches of the FOV, where there are only faint stars. An
additional difficulty is that the radius of the PSF seeing foot, as
determined from the guide star, is very large, $\sim60$\,pixels or
$1.62"$. Therefore, it is all but impossible to find bright, isolated
stars homogeneously distributed over the field that could adequately
sample the local PSF.  Nevertheless, as shown in the section above, a
reliable PSF, including its faint, very extended wings, can be
extracted from the brightest star in the field, IRS\,7, by applying an
iterative approach.  Anisoplanatic effects are caused by the AO
correction being optimized within the isoplanatic patch. The seeing
foot in AO PSFs is caused by the uncorrected light.

  Anisoplanatic effects will be much stronger in the core of the PSF
  than in the seeing foot. We illustrate this in Fig.\,\ref{Fig:PSFs},
  where we show the PSF at the position of the guide star and the PSF
  at a distance greater than the isoplanatic angle. \footnote {Note
    that the latter is actually a model PSF, created by convolution of
    the guide star PSF with the local kernel that was extracted from
    the Lucy-Richardson deconvolved image (see
    section\,\ref{sec:deconv} and Fig.\,\ref{Fig:deconv}).}  As can be
  seen in the difference image at the right panel in
  Fig.\,\ref{Fig:PSFs}, the difference between the two PSFs is
  negligible in the wings. Convolution with an elongated kernel of a
  few pixel size will not have strong effects on features with
  spatial scales of several tens of pixels.

  Consequently, a viable approach to local PSF fitting in an
  anisoplanatic image with inadequately distributed reference stars, as
  in the case of the presented data, may be to determine the
  cores of the PSFs locally and then merge these cores with the
  accurately determined wings of the PSF from the guide star. To merge
  the local PSFs with the guide star PSF, I used the {\it StarFinder}
  routine to repair saturated stars. After some experimenting, I
  decided to limit the locally determined core to 3 times the FWHM of
  the stars. Of course, the larger the local core size that can be chosen,
  the better, but there is a trade-off because in some areas of the
  field only faint ($Ks\gtrsim14$) stars are available for PSF
  extraction.

  Local PSF fitting was performed by dividing the image into
  rectangular fields smaller than the isoplanatic angle \citep[see
  also][]{Diolaiti:2000rz,Schodel:2007tw,Schodel:2009zr}. The
  $1024\times1024$\,pixel$^{2}$ field field was partitioned into
  $13\times13$ overlapping subimages of $256\times256$\,pixel$^{2}$
  ($\sim6.9"\times6.9"$). The shifts between subimages are thus just
  64\,pixels in the x- and/or y-direction and there is large overlap
  between the subframes.  This allows us to measure most stars a
  multiple number of times with different sets of PSFs.  Of the order
  200 PSF reference stars distributed over the full FOV were marked
  manually (using the {\it StarFinder} widget interface) before
  running the automated analysis of the subframes.

When choosing the size of the subframes there is some trade-off
between having a sufficient number of reference stars in the field for
accurate PSF estimation and the requirement to keep the subframe as
small as possible to keep anisoplanatic effects to a
minimum. The subframe size used in this work was found after
experimenting with various sizes. It is difficult to find an objective
measure for this quantity. Nevertheless, the experiments showed that
there is a significant tolerance in the results in terms of the exact
frame-size chosen, which can vary by several arcseconds.

The positions and fluxes of each star as well as the corresponding
{\it formal} uncertainties were computed by taking the mean of the
multiple measurements and the corresponding formal uncertainties from
overlapping frames (by {\it formal uncertainty} we refer in this work
to the uncertainty estimated by the {\it StarFinder} algorithm for
each fit, based on the given PSF and Gaussian and photon noise). The
astrometric and photometric uncertainties caused by the uncertainty
in the estimated PSF were estimated from the standard deviation of the
multiple measurements from the overlapping frames (We refer to this
source of uncertainty in this work as {\it PSF uncertainty}.). Since
the PSFs of the different subframes are not strictly statistically
independent (common PSF reference stars in overlapping subframes), the
uncertainty in the mean was computed by dividing the standard
deviation by $\sqrt{(N/3)}$ instead of $\sqrt{(N)}$, where $N$ is the
number of measurements for a given star.  The factor $3$ here is not
strictly mathematically derived, but estimated, motivated by the shift
between subframes being one quarter of their width. Additionally,
stars near the edge of a sub-frame (half a shift-width) are excluded
from the measurements (except near the edge of the combined FOV) to
avoid edge effects.  This means that about $1/3$ of the PSF reference
stars will be different in the shifted frame. No weighting was applied
to the PSF reference stars. Experiments with weighting had poorer
results. This may be caused by sporadic non-ideal PSF reference stars
(e.g., with very close companions). Weighting would also mean that one
bright star can dominate entire sub-frames, which counteracts the
attempt to sample the PSF as locally as possible and would 
reduce the independence of the measurements.  A comparison between the
resulting PSF uncertainties and formal photometric uncertainties is
shown in Fig.\,\ref{Fig:dphotloc}. There are two important
observations from the comparison of these two sources of uncertainty:
the formal uncertainty is a strong function of magnitude and the PSF
uncertainty appears to be constant, which ultimately limits the photometric
accuracy for bright stars.

Finally, the residuals (image minus detected-point-sources minus
diffuse emission) of the various sub-frames were combined to obtain
the residuals of the entire FOV. It is shown in the left panel of
Fig.\,\ref{Fig:residlinloc}.  The residuals are reduced significantly 
compared to the single PSF case. There are some negative residuals,
which are especially visible around the brightest sources. The reason for these
residuals is the difficulty of local PSF extraction combined with the
difficulty of merging the local PSF core with the wings of the PSF
extracted from IRS\,7. It may be possible to improve this process, but
the related systematic errors are not larger than a few percent (see
section\,\ref{sec:simulation}, where this method is tested on a
simulated image).

The next sub-section describes a way in which the local PSF fitting can be
improved by combining it with linear deconvolution techniques.

\subsection{Deconvolution assisted local PSF fitting \label{sec:deconv}}

\begin{figure}[!htb]
\includegraphics[width=\columnwidth]{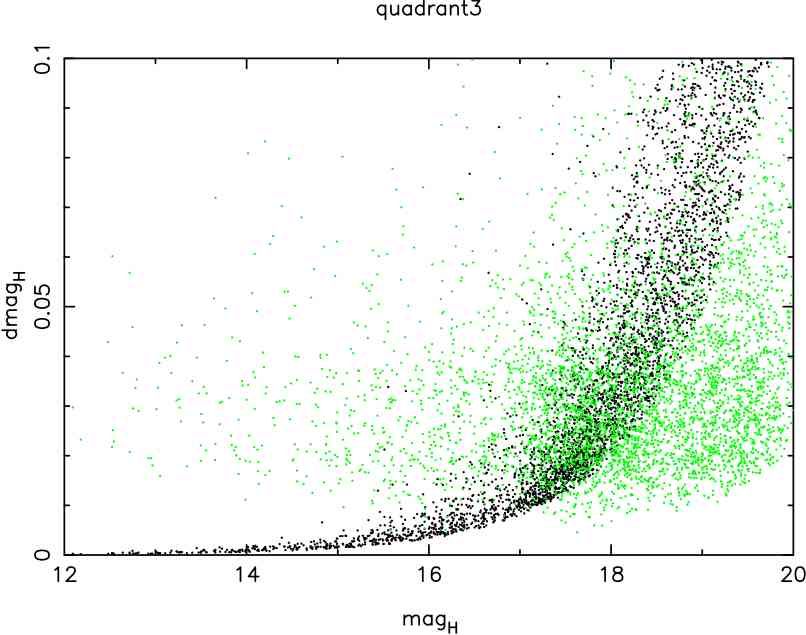}
\caption{\label{Fig:dphotlinloc} Uncertainty in PSF fitting photometry
  applied to the image of dither position 3, like in
  Fig.\,\ref{Fig:dphotloc}. The difference here is that the image was
  deconvolved with the guide star PSF and the linear Wiener-filter
  algorithm prior to local PSF fitting.
}
\end{figure}

An astronomical image can be described by a convolution of the target
($\delta$-functions in case of point-like sources) with the PSF:
\begin{equation}
f(x,y) = o(x,y)\otimes p(x',y')
\end{equation}
The target, $o(x,y)$, contains all the information about the observed
object, while the PSF, $p(x',y)$, describes the imaging process
(e.g., influence of atmosphere and telescope transfer functions). If
the PSF is known, the object can be reconstructed by deconvolution
techniques. Being an inverse process, deconvolution is, however,
always an ill-posed problem, mainly because of limited knowledge of
the PSF -- due to the importance of noise and the limited coverage of
spatial frequencies -- and the presence of noise in the
image. Deconvolution algorithms therefore have to use regularization
techniques (e.g., Wiener filtering) to produce well-behaved
solutions. For our work, we use and compare two common methods:
(a) the Lucy-Richardson deconvolution, a maximum likelihood solution
based on Bayes' theorem \citep{Richardson:1972vn,Lucy:1974yq}; and (b)
linear Wiener filter deconvolution, where the image is divided by the
PSF in Fourier space and a Wiener filter regularizes the solution
(see, e.g., \citealt {Starck:2006kx} for a detailed description of the
deconvolution problem and common methods).

We show a beam-restored (with a Gaussian beam of 3 pixel FWHM)
Lucy-Richardson (LR) deconvolved image of the FOV from dither
position~3 ($H$-band) in the left panel of Fig.\,\ref{Fig:deconv} and
the same image deconvolved with a linear Wiener filter (LW) method in
the right panel of this figure. The same PSF, extracted from IRS\,7,
was used in both cases (see inset in left panel of
Fig.\,\ref{Fig:psfextract}). I used the implementations of the
LR algorithm from the astronomical image processing package
{\it dpuser}, developed originally by \citep{Eckart:1991nx} and
substantially extended and maintained by Thomas Ott \footnote{{\it
  http://www.mpe.mpg.de/$\sim$ott/dpuser/index.html}}. The Wiener
deconvolution was programmed using the  IDL programming language.

In the presence of anisoplanatic effects, the PSF varies across the
FOV. This can be described by a convolution with a spatially variable
kernel:
\begin{equation}
PSF(x,y) = p(x,y)\otimes K(x',y'),
\end{equation}
where $p(x,y)$ is the PSF at the position of the guide star. Hence,
when we deconvolve an AO image with the guide star PSF, we are left
with $\delta$-functions at the positions of point sources in the ideal
and isoplanatic case and with functions that describe the local kernel
in the anisoplanatic case. Here, it must be noted that because of the
discrete sampling of the image, a star is practically never described
by a $\delta$-function. This would only be the case for a perfectly
known PSF, at the location of the guide star, and if the stellar
position happened to fall exactly onto the center of a
pixel. Therefore, stars are always described by local kernels
convolved with the PSF. Examples of local kernels are illustrated by
the small insets in the left image in Fig.\,\ref{Fig:deconv}. If the
guide star PSF has been used for deconvolution as in the example
presented here, then the most compact kernels are found near the guide
star, while the kernel is considerably more complex at distances
larger than the isoplanatic angle.

How can deconvolution help to improve photometry in the presence of
anisoplanatic effects? As we have seen, local PSF extraction is
necessary in the presence of anisoplanacy. The main problem with local
PSF extraction, however,  occurs in obtaining accurate estimates of the wings of the
PSFs. After deconvolution of the image with the guide star PSF, stars
in the FOV appear in the shape of the local kernels at the
corresponding positions. The sizes of these kernels are considerably
smaller than the size of the original PSF. Hence, crowding is reduced.
If we now perform the local PSF fitting algorithm, which was described in
the preceding section, on the devonvolved image, the stars can be
fit with the local kernels. Because of the reduced crowding and the
increased S/N of the point sources in the deconvolved image,
the estimation of the local kernels can be achieved with high accuracy.  In this
way, we can realize local PSF fitting and circumvent the problem of
having to truncate the locally extracted PSFs. This method will
therefore lead to improved photometry of both the point sources and the
diffuse background light.

The photometric uncertainties in the local PSF fitting photometry for the
image of dither position~3 after prior LW deconvolution (using the
guide star PSF) are shown in Fig.\,\ref{Fig:dphotlinloc}. When
comparing with Fig.\,\ref{Fig:dphotloc}, one can see that the LW
deconvolution prior to local PSF fitting leads to reduced scatter and
generally lower PSF uncertainty. The formal uncertainties, on the other
hand, appear to be slightly increased (they have been scaled by a
factor of $3$, see section\,\ref{sec:simulation}).  The residuals
related to point sources are not extended and appear to be homogeneous
across the FOV (right panel of Fig.\,\ref{Fig:residlinloc}).

We note that deconvolution violates to a certain degree a basic
assumption of PSF fitting, which is that the noise for each pixel is
independent of that of adjacent pixels. Deconvolution will lead inevitably to
covariances between the pixels.  A variety of tests show that linear
deconvolution does not lead to any significant bias, but care must be
taken when assessing the uncertainties in the measured quantities
adequately.  This issue will be discussed in section\,\ref{sec:noise}.

In the following subsection, we examine the effects of
deconvolution by working on artificial images. We study the
question of which deconvolution technique (LW or LR) is most closely suited
to our purpose.


\section{Comparing methods on a simulated image \label{sec:simulation}}

\begin{figure}[!htb]
\includegraphics[width=\columnwidth]{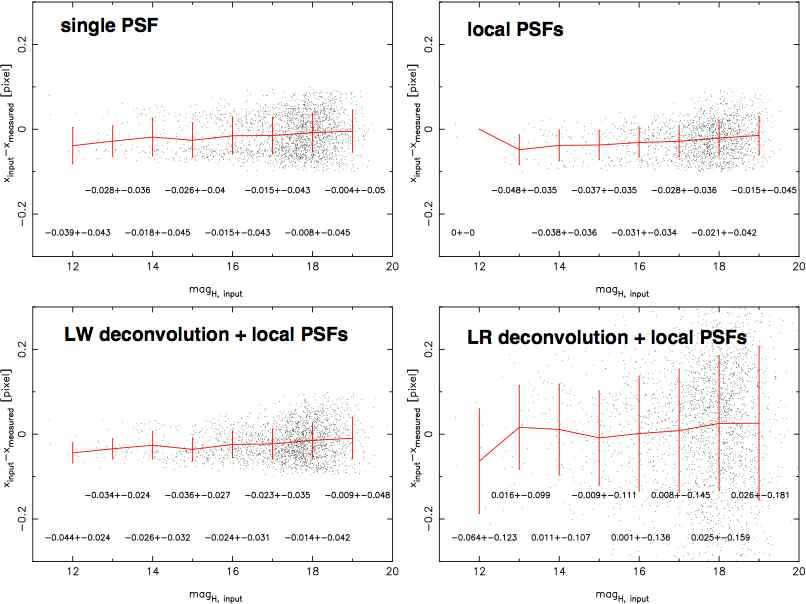}
\caption{\label{Fig:astrox} Comparison of input and measured x-axis
  positions in the simulated image. The labels give the mean
  differences and standard deviations (in pixels) per magnitude
  interval. }
\end{figure}

\begin{figure}[!htb]
\includegraphics[width=\columnwidth]{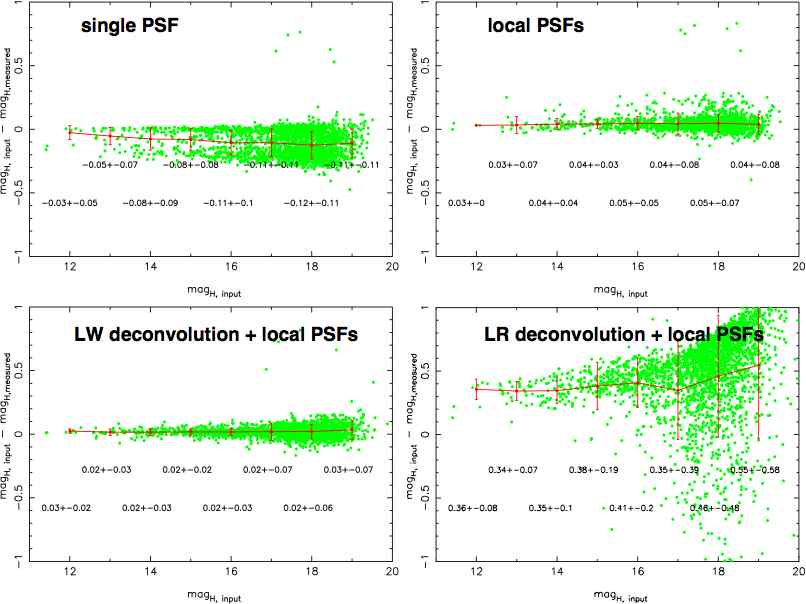}
\caption{\label{Fig:photo} Comparison of input and measured point
  source magnitudes in the simulated image.  Green dots are the
  differences between input and output values (shifted along the
  y-axis). The labels give the mean differences and standard
  deviations per magnitude interval.}
\end{figure}

In order to compare the performance of the various methods of
photometry described above we created a simulated image. The
anisoplanatic effect was modeled by using local PSFs to create the
artificial image. For this purpose, the local kernels extracted
  from the LR deconvolved image from dither position 3 (see left panel
  in Fig.\,\ref{Fig:deconv}) were used.  The kernels were extracted in
  a grid pattern from $256\times256$\,pixel$^{2}$ subframes separated
  by steps of 128\,pixels.  An individual kernel was produced for each
  source by interpolating the kernels from the four grid points
  closest to the source, except for sources at the edge of the field,
  where the nearest kernel was used, without interpolation.
  Subsequently, each kernel was convolved with the PSF extracted from
  the guide star, IRS\,7, and added to the artificial image at given
  positions and with a given flux.  In this way, an image with a
  smoothly varying PSF was created. \footnote{Note that the PSFs
    created in this way will be slightly broader than the PSFs in the
    original image because of the non-ideal properties of the kernels,
    as described in section\,\ref{sec:deconv} above. Broader PSFs will
    be a more challenging and therefore conservative way of testing the
    methods.} The diffuse emission was set to a constant value. The
fluxes and positions of the point sources were taken from the image of
dither position 3.  Gaussian readout noise and Poisson noise were
  added and the number of averaged exposures was chosen to coincide
  with the corresponding values of the data (see
  Table\,\ref{Tab:Obs}). PSF fitting photometry was performed in four
different ways: (a) by extracting the PSF from the guide star, IRS\,7, and fitting
the entire image with this single PSF; (b) marking $\sim200$ reference
stars over the entire FOV, local PSF fitting by creating local
  PSFs via extraction of  PSF cores in overlapping subframes of size
  $\sim6.9"\times6.9"$, followed by merging with PSF wings from guide
  star, as described in section\,\ref{sec:spatial}; (c) Wiener
deconvolution with the PSF extracted from IRS\,7 followed by local
extraction and fitting of PSFs; (d) as (c), but using LR
deconvolution.

\begin{figure*}[!htb]
\includegraphics[width=\textwidth]{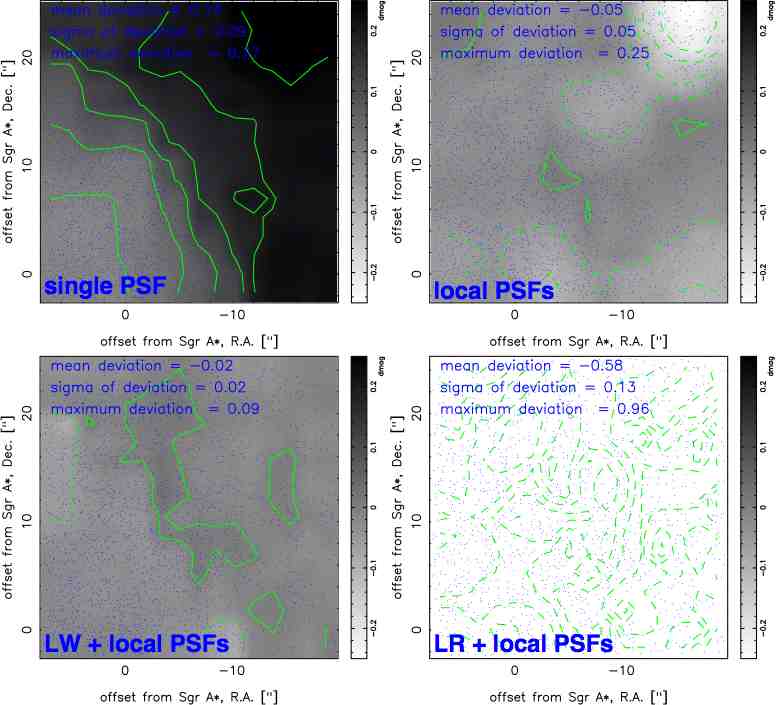}
\caption{\label{Fig:dmag} Smoothed maps of the median differences (in
  boxes of $64\times64$ pixels) between the measured magnitudes of
  stars in a simulated image and the input magnitudes. The dots are
  the locations of detected stars. Upper left: PSF fitting with a
  single, fixed PSF extracted from the guide star. The contour lines
  indicate differences of $0.0-0.25$ magnitudes in steps of
  $0.05$. Upper right: PSF fitting with locally extracted
  PSFs. Contour lines are plotted at $0.05$ to $-0.2$ magnitudes in
  steps of $0.05$. Lower left: PSF fitting with locally extracted PSFs
  after LW deconvolution with the guide star PSF. Contour lines are
  plotted at $-0.05$ (dashed), $0.0$ , and $0.05$ magnitudes.  Lower
  right: PSF fitting with locally extracted PSFs after LR
  deconvolution with the guide star PSF. Contour lines are plotted at
  $-0.3$ to $-0.9$ magnitudes in steps of $0.1$. Note that in this
  case the panel 
  is off the gray scale, which has been kept constant for all four panels.}
\end{figure*}

\begin{figure*}[!htb]
\includegraphics[width=\textwidth]{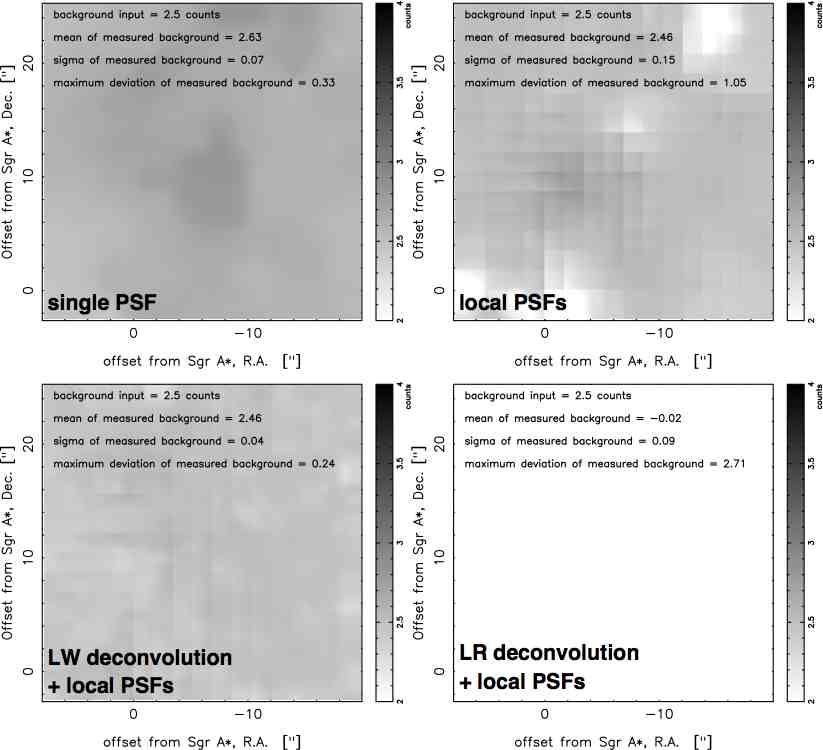}
\caption{\label{Fig:backsim} Smooth background emission estimated for
  the point-source subtracted simulated image using different PSF
  fitting methods. Upper left: single,
  constant PSF. Upper right: local PSFs. Lower left: linear Wiener
  filter deconvolution and local PSF fitting. Lower right:
  Lucy-Richardson deconvolution and local PSF fitting. The gray scale
  is linear. Note the different gray scale in the lower right panel.
  The input background was constant and set to $2.5$\,counts
  (corresponding to a surface brightness of
  $15.5$\,mag\,arcsec$^{-2}$). The labels give the mean difference,
  standard deviation, and maximum absolute deviation between measured
  and input background.}
\end{figure*}

The astrometry and photometry of the recovered point sources was
finally compared with the input values. The extracted smooth diffuse
background was compared to the input background (chosen to be a
constant). The results are illustrated in
Figs.\,\ref{Fig:astrox}, \ref{Fig:photo}, \ref{Fig:dmag}, and \ref{Fig:backsim}.

\subsection{Astrometry}

The differences between input and recovered positions for the
different methods are shown in Fig.\,\ref{Fig:astrox} (only the x-axis
values are shown, the y-axes values showing very similar behavior).
LW deconvolution with subsequent local PSF fitting allows us to
recover the positions of point sources with a standard deviation
$<0.04$\, pixels, except for the faintest sources. Local PSF fitting
without deconvolution and fitting with a single PSF lead to results
that are of similar quality. Lucy-Richardson deconvolution clearly
deteriorates the astrometry, the standard deviations of the stellar
positions being 2-3 times higher than in the other cases. We note that
the mean of the differences between input and recovered positions is
different from zero in all cases (by a few $1/100$ of a pixel). I
propose that the most important factor for this behavior is that the
position of a star has been defined in the simulated images to
coincide with the centroid position of the PSF (this is the usual
standard for PSF fitting algorithms). In the artificial image, the
stars are fixed at their locations and convolved with the complete
PSF. StarFinder (like DAOPHOT) fits the positions and fluxes of the
sources only within a fitting radius (or a fitting box in the case of
StarFinder) and uses the complete PSF only for point source
subtraction. The centroid of the PSF within the fitting radius (box)
will differ from that of the entire PSF if the PSF outside the fitting
radius is not point-symmetrical. That is, the difference in mean
positions between input and measurement is due to a difference between
the PSFs used for input (entire model PSF) and output measurement (a
partial PSF or a deconvolved PSF). In the case of a single PSF, it
appears that there are actually {\it two} distributions of the
differences between input and output, one centered on a slightly
positive position, another one centered on slightly negative
values. This is not necessarily surprising because the PSF becomes
elongated with distance from the guide star, which may change the
centroid position, depending on which side of the guide star a star is
located.  However, I did not fully explore this possibility because it
would go far beyond the scope of this work if I were to explore the
phenomenon of deviation between input and measured positions in detail
here. However, I believe it is important to mention these points here
because they may become of great significance in work that requires
extremely precise astrometry in AO images. It may not be sufficient to
determine stellar positions just from the PSF cores. At the moment,
also I cannot exclude the possibility that positions may become biased
by deconvolution, depending on the location in the field.

\subsection{Photometry}

The differences between input and recovered positions for the
different methods are shown in Fig.\,\ref{Fig:photo}. Both PSF fitting
with a single PSF and PSF fitting after LR deconvolution produce
significant deviations with the latter method providing the poorer
results. An explanation of the bad performance of the LR algorithm
can probably be found in its non-linearity. The LR deconvolution tends
to be influenced by local noise peaks and incorporates the smooth
background into the point sources. This leads to an increasing
overestimation of the flux of faint sources and to characteristic
empty patches with a size similar to the PSF around bright sources.
Both local PSF fitting and local PSF fitting after linear
deconvolution provide acceptable results. Local PSF fitting after
Wiener deconvolution leads to the smallest standard and 
mean deviations. We note that the distribution for the single
PSF (upper right panel in Fig.\,\ref{Fig:photo}) appears
bivariate. This may be related to the PSF properties discussed in the
previous paragraph. Positive deviations, i.e., a source that is
brighter in the measurement than in the input, are largely excluded in
the case of a single PSF used for measurement. Elongation of the sources
with distance from the guide star will lead to a loss of flux.

Smooth maps of the differences between input and measured photometry
across the FOV are shown in Fig.\,\ref{Fig:dmag}. Using a single PSF
leads to a systematic error that increases with distance from the
guide star up to $\sim0.25$\,mag. LR deconvolution combined with local
PSF fitting does not lead to acceptable results, as we have already
seen in Fig.\,\ref{Fig:photo}. Local PSF fitting works quite well, 
only leading to systematic deviations in the upper right corner of the
image, at extreme distances from the guide stars, where the PSFs are
notably elongated. As an additional difficulty, there are no bright,
isolated stars available in this region. Again, linear deconvolution
followed by local PSF fitting is the most reliable method, leading to
the smallest systematic variations across the FOV.

\subsection{Diffuse background}

The diffuse background emission in the artificial images was set to
$2.5$\,counts, corresponding to $15.5$\,mag\,arcsec$^{-2}$. To ensure
a reliable extraction of the background emission, it is again the
Wiener deconvolution combined with local PSF fitting that produces the
best results (Fig.\,\ref{Fig:backsim}). While single PSF fitting and local PSF fitting without
deconvolution also produce acceptable results, they nevertheless cause
significant systematic deviations across the FOV. LR deconvolution has
the tendency to ``scoop up'' the diffuse emission into the point
sources. The information on the background is largely lost in the LR
deconvolved image. The smooth background emission is instead
incorporated into the point sources.  This is why the magnitudes of
point sources in the LR deconvolved image become increasingly biased
the weaker the sources are (see bottom right panel of
Fig.\,\ref{Fig:photo}) \footnote{This effect was taken care of
  in \citet{Buchholz:2009sp}, whose method assured accurate {\it relative}
  calibration of the point sources.}.

We conclude that the optimal method -- among the ones that have been
considered in this work -- for extracting reliable photometry (and
astrometry) for adaptive optics images with a FOV larger
than the isoplanatic patch and sparsely sampled PSF  is based on the following steps: 
\begin{enumerate}
\item PSF extraction from the guide
star (or from (a) bright star(s) near the guide star).
\item  Wiener deconvolution of the image.
\item Local PSF fitting. 
\end{enumerate}
We note that this method is not restricted to the case of the GC. It can
probably be applied successfully in other cases where one is
interested in accurate photometry over a large FOV in AO observations
of crowded fields, but has to deal with sparse sampling of the PSF.

\section{Further tests}

\subsection{Deconvolution, noise, and uncertainties \label{sec:noise}}

  Deconvolution - at least in the linear case -
  can be regarded as  re-imaging the data with a different set of
  (virtual) optics.  However, the difference from real optics is that
  the true image has already been made and that the noise is already
  present in the data, the latter point explaining why deconvolution 
  needs to be applied with some care. In this section, I discuss
  whether it is valid to combine deconvolution with PSF fitting
  techniques and what caveats have to be kept in mind.

\begin{figure}[!htb]
\includegraphics[width=\columnwidth]{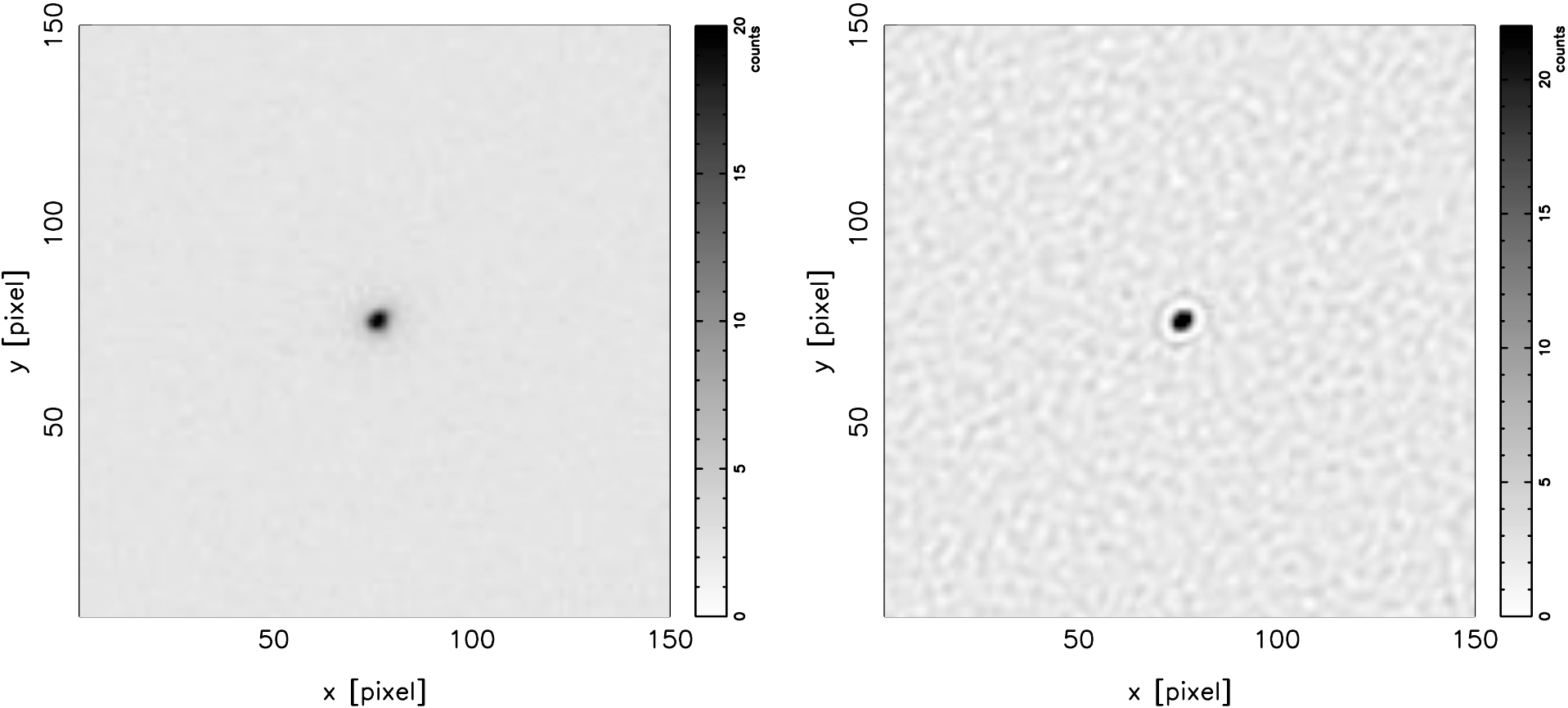}
\caption{\label{Fig:artim} Simulated image, using a background count
  of $2.5$, readout noise of $4.2$, electrons-per-ADU of $11.0$,
  corresponding to the used NACO mode (224 exposures were used, i.e.,
  $N=8$, $NDIT=28$). The single source at the center has
  1000~counts. Left: raw image. Right: Wiener deconvolved image.  The
  standard deviation of the raw sky is $0.34$ counts, the one of the
  deconvolved sky $1.49$ counts. The mean of the two skies is, of
  course, identical.}
\end{figure}

  PSF fitting is supposed to be applied to raw images, i.e., the noise
  statistics of the pixels should be preserved, which will be used to
  assign weights to the individual pixels
  \citep[e.g.,][]{Stetson:1987nx}. Deconvolution, even if it is linear
  as in Wiener deconvolution, will violate this assumption to a
  certain degree because it will lead to covariances between the
  pixels. This is illustrated in Fig.\,\ref{Fig:artim}, which shows
  the raw and the deconvolved version of an artificial
  image of a sky ($2.5$ counts) containing 1 star (1000 counts): The
  deconvolved sky shows some ``granularity'' caused by the covariances
  between the pixels.  \footnote {Note that in this work the Gaussian
    part of the noise in the data was estimated directly from the
    image and not from a number of exposures and readout noise. This is
     done by the {\it StarFinder} routine
    GAUSSIAN\_NOISE\_STD. Although this does not take care of the
    covariances, in this way it is at least possible to use a more
    conservative estimate of the noise due to Gaussian processes.}  In
  a Monte Carlo simulation, 100 realizations of this image plus star
  were created. Subsequently, the position and flux of the star as
  measured in the raw and deconvolved images were
  compared. Table\,\ref{Tab:artstat} gives the mean of the recovered
  position and flux as well as their corresponding standard deviations
  and the mean of the formal uncertainties estimated by
  {\it StarFinder}.  The result of this simulation shows that the
  position and flux of the star are reliably recovered from the
  deconvolved image. However, the standard deviation of the
  measurements for the deconvolved image is somewhat larger than in the
  raw image (e.g., increase in flux uncertainty from $0.4\%$ in the
  raw image to $0.7\%$ in the deconvolved image) and the PSF fitting
  algorithm underestimates the true uncertainties of position and
  flux by a factor of $\sim3.3$.

\begin{table*}
\centering
\caption{Results of Monte Carlo modeling of
  background plus one star (100 tries) as shown in
  Fig.\,\ref{Fig:artim}.}
\label{Tab:artstat} 
\begin{tabular}[htb]{l|lllllllll}
 & $x^{\mathrm{a}}$ & $\sigma_{\rm x,formal}^{\mathrm{b}}$ & $\sigma_{\rm x,true}^{\mathrm{c}}$  & $y^{\mathrm{d}}$ & $\sigma_{\rm y,formal}^{\mathrm{e}}$ & $\sigma_{\rm y,true}^{\mathrm{f}}$ & $f^{\mathrm{g}}$ & $\sigma_{\rm f,formal}^{\mathrm{h}}$ & $\sigma_{\rm f,true}^{\mathrm{i}}$ \\
 & [pix] & [pix] & [pix] & [pix] & [pix] & [pix] & [counts] & [counts]  & [counts]\\
\hline
\hline
raw & 75.670 & 0.009 & 0.009 & 75.109 & 0.008 & 0.009 & 1000 & 4 &4\\
deconvolved & 75.670 & 0.003 & 0.011 & 75.108 & 0.003 & 0.010 & 1000& 2 & 7 \\
\hline
\end{tabular}
\begin{list}{}{}
\item[$^{\mathrm{a}}$] Measured position of the star on the X-axis (model position $x=75.670$)
\item[$^{\mathrm{b}}$] $1\,\sigma$ uncertainty of  measured $x$ given by the PSF fitting algorithm (average of the uncertainties of all the tries)
\item[$^{\mathrm{c}}$] $1\,\sigma$ uncertainty of measured $x$ as given by the standard deviation of the individual measurements
\item[$^{\mathrm{d}}$] Measured position of the star on the Y-axis (model position $x=75.110$)
\item[$^{\mathrm{e}}$] $1\,\sigma$ uncertainty of measured $y$ given by the PSF fitting algorithm (average of the uncertainties of all the tries)
\item[$^{\mathrm{f}}$] $1\,\sigma$ uncertainty of measured $y$ as given by the standard deviation of the individual measurements
\item[$^{\mathrm{g}}$] Measured flux of the star (input flux $f=1000.0$)
\item[$^{\mathrm{h}}$] $1\,\sigma$ uncertainty of measured $f$ given by the PSF fitting algorithm (average of the uncertainties of all the tries)
\item[$^{\mathrm{i}}$] $1\,\sigma$ uncertainty of measured $f$ as given by the standard deviation of the individual measurements
\end{list}
\end{table*}

\begin{figure*}[!htb]
\includegraphics[width=\textwidth]{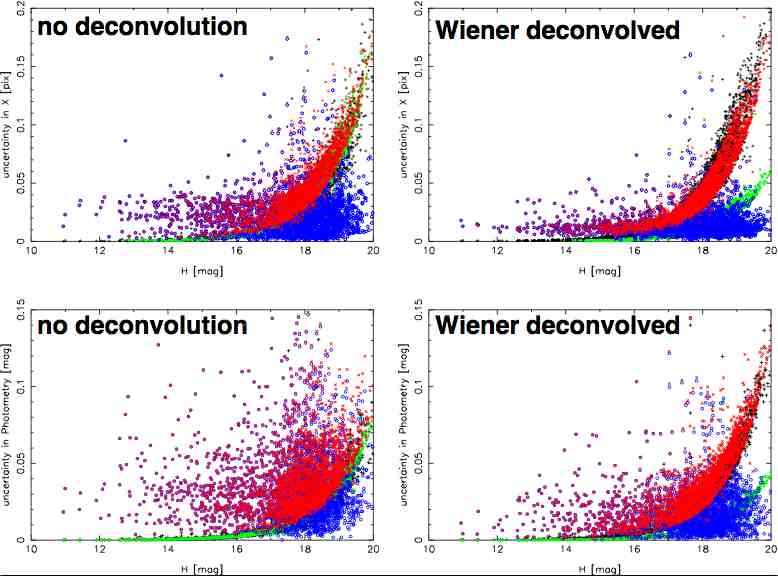}
\caption{\label{Fig:simulation} Results of Monte Carlo simulations
  (100 runs) on an artificial image. Left: no deconvolution, PSF
  fitting after determining local PSFs via combination of local cores
  with the wings of the guide star PSF. Right: Wiener deconvolution
  before local PSF fitting. Upper panels: Uncertainty in X-positions
  vs.\ magnitude. Lower panels: Uncertainties in photometry vs.\
  magnitude. Black plus signs (hardly visible in most parts, generally
  underlying the red crosses) are uncertainties determined from the
  standard deviations between input and measured positions and
  fluxes. Green stars are the formal uncertainties estimated by the
  PSF fitting algorithm. Blue rings are the PSF uncertainties. Red
  crosses are the combined formal plus PSF uncertainties. In the case
  of the Wiener deconvolved images, the formal uncertainties were
  scaled by a factor of 3 before combining them with the PSF
  uncertainties.}
\end{figure*}

In a second simulation, 100 Monte Carlo simulations were run on the
artificial star field used in section\,\ref{sec:simulation}. The local
PSF fitting algorithm was performed on both the raw and the deconvolved
images. In the case of the raw images, local PSFs were created by merging
locally extracted cores with the wings from the guide star PSFs. The
guide star PSF was used for the deconvolution. Input and recovered
positions and fluxes were compared to determine the true standard
deviations of these quantities. Those were subsequently compared with
the formal and PSF uncertainty estimates delivered by the PSF fitting
algorithm. It can be seen in Fig.\,\ref{Fig:simulation} that the
scatter in the uncertainties is lower when Wiener deconvolution is
applied. It can also be clearly seen that the formal uncertainties
estimated by the PSF fitting algorithm (green stars) are
underestimated in the deconvolved images. The correct uncertainties
can however be reproduced when the formal uncertainties in the
deconvolved images are scaled by a factor of $\sim3$ before
quadratically combining them with the PSF uncertainties. We note that the
simulations may indicate that the PSF uncertainty is
overestimated (at least for the bright stars). Since this is less of a
problem than under-estimating the uncertainty, I have not 
investigated this point further for the time being.

In addition to these simulations, several other tests were
performed to check whether the Wiener deconvolution introduces a
significant bias in the astrometry or photometry of the data used in
this work:
\begin{itemize}
\item As shown in section\,\ref{sec:simulation}, the simulated data (using numbers of
exposures, read-out noise, gain corresponding to the data) show that
Wiener deconvolution leads actually to the {\it best} astrometric and
photometric results in the
case of the presented data because it allows an accurate estimate of the
local PSF. 
\item In section\,\ref{sec:realtest}, the uncertainties obtained from  the photometry
  with Wiener deconvolution combined with local  PSF fitting is
  checked by comparing results for
  sources present in the overlapping region of all four dither
  positions (see section\,\ref{sec:data} and
  Fig.\,\ref{Fig:mosaic}). This test suggests that the applied method
  delivers an accurate estimate of the photometric uncertainties.
\item Finally, Wiener deconvolution combined with  local PSF fitting
  was applied separately to each of the 8 exposures for dither
  position\,3. Mean positions and fluxes as well as the corresponding uncertainties of the
  sources present in all the exposures were derived from these
  independent measurements. The values were compared with the results
  obtained for the combined image. Again, the results coincide
  well within the uncertainties.
\end{itemize}

There appears to be a workaround to avoid performing PSF fitting on
the deconvolved image. It consists of extracting local kernels from
the deconvolved data and subsequently reconvolving them with the PSF
that was used to deconvolve the entire image. The result would be a
local PSF that could be applied to PSF fitting in the raw image to
thus avoid the problem of altered noise statistics in the deconvolved
image. Unfortunately, this approach does not lead to satisfactory
results. Deconvolution -- regardless of the actual method used -- is
an ill-posed problem and requires some kind of regularization. The
purpose of the regularization is to suppress noise at high
frequencies. This will lead inevitably to some loss of real
information. In the case of Wiener deconvolution, the Wiener filter
suppresses the highest frequencies. The trade-off is the
characteristic ringing around point sources, or, if the reglarization
parameter is large, to broad, Gaussian-like PSFs. The effect can be
seen, e.g., in Fig.\,\ref{Fig:artim}, where the PSF of the deconvolved
source has a finite FWHM of a few pixels and is surrounded by a ring
that is typical of Wiener deconvolution. Reconvolving the
deconvolved source will therefore {\it not} recover the original
image. The PSF in the reconvolved image will in fact be too
broad. So, one cannot use the kernels extracted from deconvolved
images to recover the PSF of the original image.

To conclude, I believe that the numerous tests applied to real and
artificial data show that, although applying PSF fitting to a Wiener
deconvolved image is not rigorously mathematically correct, the
combination of Wiener deconvolution and PSF fitting works reliably and
does not introduce any significant bias into the astrometry and
photometry.  But care must be taken to arrive at correct estimates of
the uncertainties. The great advantage of (linearly) deconvolving the
data is that one is then able to extract reliable local PSFs.
However, there is a price that has to be paid. Deconvolution appears
to increase the uncertainty in the measured positions and
fluxes. Because of the covariances between the pixels that are caused
by the deconvolution process, these uncertainties cannot be estimated
directly from the PSF fitting algorithm. Instead, scaling factors must
be applied to the formal uncertainties. The good news is that the
scaling factor can be estimated in a fairly straightforward (but
time-consuming) way by artificial star tests. Alternatively, multiple
measurements of independent data sets can be combined to estimate the
uncertainties and eliminate spurious sources.

\subsection{DAOPHOT  \label{sec:daophot}}

The software package DAOPHOT, which is also included in the {\it IRAF}
program package, can be considered to represent the industrial
standard on PSF fitting. It allows us to choose between six kinds of
numerical PSFs, which are allowed to vary linearly or quadratically
across the FOV. Additionally, a look-up table is produced to take into
account the deviations of the true PSFs from the numerical
model. Since a comparison with the standard is obviously important, I
experimented on the data used in this paper with DAOPHOT, although,
unfortunately, I was unable to produce any satisfying results. While
I cannot exclude that this unsatisfying outcome is due
to my lack of experience with DAOPHOT I nevertheless believe to have
identified the following reasons for not having been successful:

\begin{itemize}
\item Peculiar properties of low Strehl AO PSF. The FWHM of the guide
  star in the data used here is $\sim3.5$\,pixel or $0.095"$. The
  seeing foot of the complete PSF, however, has a radius of
  $\sim60$\,pixels or $1.62"$. This is more than a factor of 10 larger
  than the PSF core. This behavior is found in neither seeing-limited
  data nor data from  HST, to which DAOPHOT is frequently
  applied. Additionally, there are just a handful of stars present in
  the image that are bright enough for the PSF to be traceable out to
  these large distances. For fainter stars, the PSF apparently
  disappears within the noise at radii $\sim30-40$\,pixels. Nevertheless,
  the combined action of the overlapping PSF wings of highly crowded
  bright stars, such as around Sagittarius\,A* (located at the origin
  in Fig.\,\ref{Fig:mosaic}), makes it absolutely necessary to derive
  the full PSF out to 60\,pixels.  In the contrary case, one would
  also lose information about the diffuse emission de to unresolved
  stars. The latter carries valuable information about the structure of
  the Galactic center nuclear star cluster because it is currently
  possible to resolve only a few percent of the stars in this region
  \citep[see][]{Schodel:2007tw}.
\item Crowding of the field. Because of the large size of the PSF it
  is very difficult to find sufficiently bright and isolated stars as
  PSF references.  IRS\,7 is by far the most reliable PSF reference
  and in fact the only one that allows us to reliably estimate the PSF
  wings at large distances from the star.
\item Inhomogeneous distribution of stars. To derive a
  reasonable PSF model, it is important to have a sufficient number of
  isolated stars distributed roughly homogeneously across the FOV. This
  is clearly not the case here, where there are large patches without
  bright stars. Some of the bright sources in the field are also
  extended (bow-shocks and/or relation to diffuse emission in
  the close vicinity of the star).   The lack of bright stars in some
  areas of the field, combined with the extended PSF wings, also
  appear to lead to systematic errors in the PSF look-up tables.
\item  Naturally, I experimented with de-crowding the field with purely numerical
  or with only linearly variable PSFs. Unfortunately, this produced
  large residuals and did not lead to success. Possible reasons are
  that the available PSFs do not model the seeing foot of
  the PSF  well enough in this case. An additional difficulty, which
  may only be true for the data used here is that even the core of the PSF is
  not symmetric (which is not rare in AO). An indication of the
  inadequacy of the available mathematical PSF models may be that there was a strong
  degeneracy as to which mathematical model to apply for the PSF
  (basically all available functions gave very similar chi-squared values).
\end{itemize}

All these points can be briefly summarized in the statement that the
PSF is complicated, very extended, and variable across the field, but
there is only one good PSF reference star available (3 reference stars
per degree of freedom are a minimum requirement for DAOPHOT and
similar approaches). This lies at the root of this exercise. If the
situation were easier, simple local PSF extraction with {\it
  StarFinder} would have produced satisfying results as well. I am
confident that in this case DAOPHOT (or SExtractor) would have worked
very well too. It is probably possible to modify the standard
algorithms in ways that can deal with the described difficulties. These
ways are indicated by the observation that the true seeing foot of
the AO PSF does not vary drastically with position in the
field. An accurate numerical description of the seeing foot combined with
look-up tables for just the PSF cores may be an avenue worth
exploring.

\subsection{Choice of initial PSF}

A source of uncertainty that has not yet been considered is the PSF
used for the linear Wiener-filter deconvolution before the local
point-source fitting. Because of the bright guide star, IRS\,7, it is
fairly easy to obtain an accurate estimate of the guide star PSF for
the observations analyzed here. However, it is not obvious how to
proceed in cases where there is no sufficiently bright and isolated
star available for PSF extraction, as in a crowded field full of
faint sources, or when the bright stars are saturated. Apart from
these points, one must also keep in mind that the PSF used for
deconvolution is always subject to uncertainties. How flexible is the
presented method in terms of its constraints on the PSF used for the
deconvolution?

Linear Wiener filter deconvolution is a {\it linear} process, i.e., it is
reversible -- with the caveat that some information is suppressed by
the need to cut high frequencies with the Wiener filter. If there is a
certain degree of inaccuracy in the PSF, this inaccuracy will be
conserved in the local kernels after deconvolution. Since we fit the
stars with the local kernels in the second step, any systematic
uncertainty should be largely taken care of by the suggested method. We
therefore expect no strong biases that may be introduced by an
inaccurate PSF in the first step. 

We performed tests of this hypothesis. The simulated image from
section\,\ref{sec:simulation} was LW deconvolved with different PSFs
prior to local PSF fitting. We used (a) a broadened version of the
guide star PSF (convolution of the guide star PSF with a 2 pixel
Gaussian, i.e.,s core of the PSF too broad), (b) a narrower version of
the guide star PSF (obtained by raising the PSF to the power of 1.1 at
all pixels), (c) a PSF derived from 9 mag$_{H}\approx10$ stars within
$5"$ of the guide star (i.e., under-estimation of the wings), and (d)
a Moffat function that was fitted to the guide-star PSF (errors in both
core and wings of the PSF). The systematic errors in the photometry of
the point sources as a function of position in the FOV were limited to
$\lesssim0.03$\,mag in all cases. Concerning deviations and
uncertainties in photometry and astrometry, as well as the reliability
to recover the diffuse background emission, the results are summarized
in Table\,\ref{Tab:psferr}. The smallest errors correspond to when the errors
of the PSF are confined to its core, while errors in the wings will
have a more significant influence on the photometry, particularly on
the reliability of the recovered diffuse emission. Not surprisingly,
the pure mathematical model, the Moffat function, produces the poorest -
but still satisfying - results (errors in both the core and wings of
the PSF).

\begin{table}[!htb]
\centering
\caption{Overview of consequences of errors in the estimated PSF.}
 \label{Tab:psferr}
\begin{tabular}{llllll}
  &  $\Delta_{\rm mag}^{\mathrm{e}}$ & $\sigma_{\rm mag}^{\mathrm{f}}$ & $\Delta_{\rm pos}^{\mathrm{g}}$ &  $\sigma_{\rm pos}^{\mathrm{h}}$ & $Diff_{\rm mean, \sigma, max}^{\mathrm{i}}$\\
  & [mag] & [mag] & [pix] & [pix] & [counts]\\
\hline
\hline
a$^{\mathrm{a}}$ & 0.02 & 0.02-0.11 & 0.05 & 0.03-0.1 & 2.45, 0.50, 0.4\\
b$^{\mathrm{b}}$ & 0.03 & 0.01-0.13 & 0.02 & 0.02-0.10 & 2.70, 0.26, 1.9\\
c$^{\mathrm{c}}$ & 0.04 & 0.02-0.13 & 0.05 & 0.025-0.10 & 2.34, 0.12, 1.0\\
d$^{\mathrm{d}}$ & 0.075 & 0.01-0.12 & 0.03 & 0.016-0.10 & 2.12, 0.27, 1.68 \\
\hline
\end{tabular}
\begin{list}{}{}
\item[$^{\mathrm{a}}$] PSF convolved with a 2 pixel FWHM Gaussian, i.e., too broad (mainly in the core)
\item[$^{\mathrm{b}}$] PSF with all pixels raised to the power of $1.1$, i.e., too narrow
\item[$^{\mathrm{c}}$] PSF derived from several stars within $5"$ of the guide star, but ${\sim4}$\,mag fainter, i.e. PSF  wings underestimated
\item[$^{\mathrm{d}}$] Moffat function (fitted to guide star PSF) used as PSF
\item[$^{\mathrm{e}}$] Approximate mean offset between input and measured magnitudes of the stars
\item[$^{\mathrm{f}}$] Range (for magnitudes 12 to 19) of the standard deviation of the difference between input and measured magnitudes
\item[$^{\mathrm{g}}$] Approximate mean offset between input and measured pixel positions of the stars (given for X, similar for Y)
\item[$^{\mathrm{h}}$] Range  (for magnitudes 12 to 19) of the standard deviation of the difference between input and positions (given for X, similar for  Y)
\item[$^{\mathrm{i}}$] Mean, sigma and maximum deviation of estimated diffuse background
\end{list}
\end{table}

Of course, even when using IRS\,7 as a PSF reference, there is a certain
subjectivity in adjusting the smoothing parameters for the wings (the
HALO\_SMOOTH routine) and when choosing the final masking
radius. However, several tests (varying the values of the smoothing
parameters and masking radius by up to $20\%$) show that the bias of
these effects on the point source photometry is smaller than $1\%$. It
can, however, have an effect of order $10\%-20\%$ on the estimated diffuse emission
near bright sources or in very densely populated areas.

We conclude that uncertainty in the PSF used for the primary
deconvolution will lead to uncertainties of only a few percent in the
measured flux of point sources, but can become of greater significance
for the diffuse background. While the results show some robustness
with respect to PSF errors, great care should be taken, nevertheless,  when extracting
the primary PSF, particularly its extended wings. This process is unfortunately never free of
subjectivity and requires some  experience. It is always recommendable
to check the residuals and the diffuse flux related to the area of the
image where the PSF reference star(s) is (are) located.

\subsection{Accuracy of the method tested on real data \label{sec:realtest}}

\begin{figure}[!htb]
\includegraphics[width=\columnwidth]{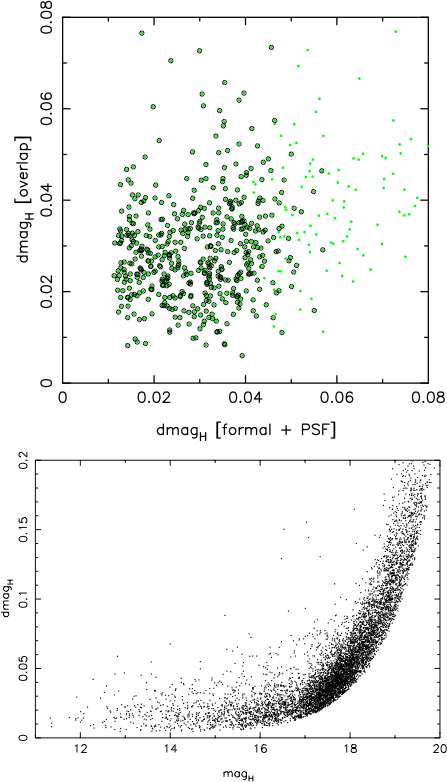}
\caption{\label{Fig:dphotcheck} Top: Comparison between photometric
  uncertainties in the $H$-band measurements obtained from the
  algorithm applied here (scaled (see section\,\ref{sec:simulation} ) formal fitting plus PSF uncertainties,
  $dmag_{formal+PSF}$) and the uncertainties derived from stars in the
  field where the four dither positions overlap
  ($dmag_{overlap}$). Green dots: all stars in the common area; black
  circles: only stars with mag$_{H}\leq18$ Bottom: Plot of photometric
uncertainty vs.\ magnitude for all stars detected in the entire FOV
(all four dither positions).}
\end{figure}

As an independent check of the photometric accuracy, we compared the
photometric uncertainties derived from our algorithm (i.e., formal plus
PSF uncertainties) with the uncertainties derived by comparing the measurement
of stars present in all four dither positions of the $H$-band observations.

We extracted a PSF from the guide star, IRS\,7, for each of the images
corresponding to the four dither positions. The images were then
deconvolved with the linear Wiener filter algorithm. Finally, local
PSF fitting was performed on the deconvolved images. A scaling factor of
$3.3$ was applied to the formal uncertainties given by the PSF fitting
algorithm to account for the under-estimation of
uncertainties in deconvolved images (see section\,\ref{sec:noise}). The lists of detected
point sources in the overlapping subframes and the smooth background
estimates and residuals for the overlapping subframes were combined
as described in section\,\ref{sec:spatial}. Uncertainties were
calculated by quadratically combining the formal fit uncertainties
with the PSF uncertainties. For stars without multiple
measurements. we adopted a PSF uncertainty of $0.02$\,mag (see
Fig.\,\ref{Fig:dphotlinloc}).

The uncertainties derived from the algorithm (formal plus PSF
uncertainties) are compared with those obtained from
the four independent measurements in the overlapping fields in
Fig.\,\ref{Fig:dphotcheck}. The uncertainties appear to be uncorrelated
and of similar magnitude. More than $91\%$ ($50\%$) of the stars have
a photometric uncertainty smaller than $0.05$\,mag ($0.03$\,mag). The
bottom panel of Fig.\,\ref{Fig:dphotcheck} shows a plot of the
photometric uncertainty versus magnitude for all stars detected in the
combined (i.e., all four dither positions) FOV. To exclude
spurious sources that may possibly originate in the deconvolved images, we
excluded stars that are not also detected by local PSF fitting
without prior deconvolution.

\begin{figure}[!tbh]
\includegraphics[width=\columnwidth]{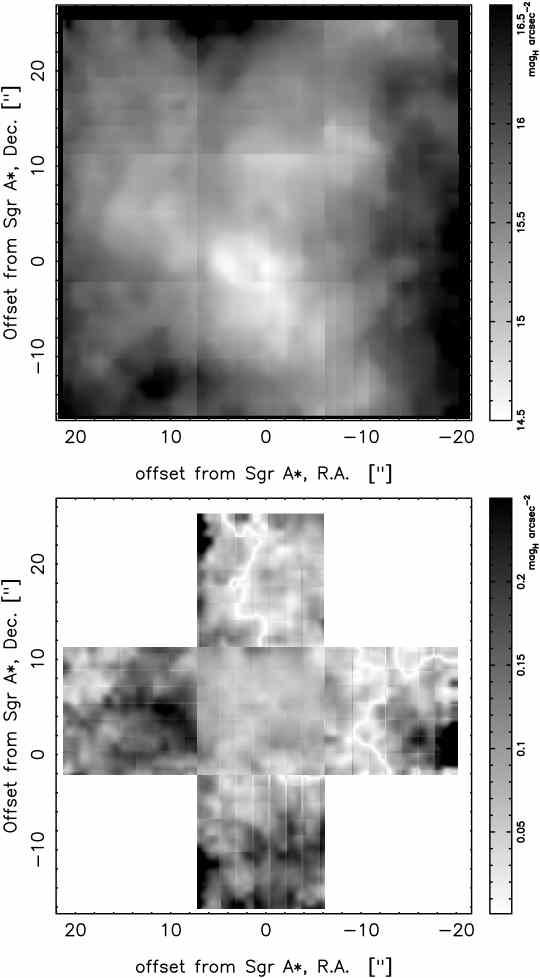}
\caption{\label{Fig:back} Diffuse emission in the $H$-band FOV. The bottom panel illustrates the uncertainty in the diffuse
  emission, estimated from the overlap areas between the dither positions. }
\end{figure}

The diffuse emission extracted form the entire FOV of the $H-$band
observations is shown in the top panel of Fig.\,\ref{Fig:back}. The
checkerboard pattern is caused by our method because we have partitioned
the field into many small overlapping sub-fields. The bottom panel of
Fig.\,\ref{Fig:back} shows the uncertainty in the measured diffuse
background determined from the deviation between overlapping
fields. It can again be seen that the applied algorithm appears to
work very well. No systematic variations can be seen and the
uncertainty is generally $\leq 0.1$\,mag\,arcsec$^{-2}$, with the
exception of some small patches, where the uncertainty can reach
$\sim 0.25$\,mag\,arcsec$^{-2}$.

\section{Summary and conclusions}

Photometry with PSF fitting algorithms depends critically on the
  ability to obtain accurate estimates of the PSF and its variability
  over the FOV. Standard software (DAOPHOT, SExtractor, StarFinder)
  can deal well with variable PSFs if there is a sufficient number of
  bright, unsaturated, and isolated stars present in the
  image. Additionally, these stars must be distributed in a roughly
  homogeneous way to adequately sample the PSF
  variability. In this work, we present a data set, an adaptive optics
  image of the Galactic center, that presents several particularities
  that make a standard approach difficult, leading to large systematic
  errors. The main difficulties are extremely extended PSF wings
  combined with strong crowding, significant variability of the PSF
  due to anisoplanatic effects, and, above all, a scarcity of adequate
  PSF reference stars in most parts of the fields. There is
  only a single excellent reference star present in the images.  Local
  extraction of PSFs is limited in its accuracy by the variable
  density and brightness of the stars in the field, which leads
  generally to an underestimation of the extended wings of the PSFs
  and therefore an under-estimation of the brightness of point sources
  and a corresponding overestimation of the diffuse emission.

  This work shows that in such a difficult case one can obtain
  accurate photometric and astrometric results using two methods: (a)
  assuming that the wings of the PSFs, i.e., the seeing foot of the AO
  PSF, do not vary significantly across the FOV. In that case, local
  PSFs can be estimated by combining locally extracted PSF cores with
  the wings of the PSF of a bright star (e.g., the guide star); (b)
  using the PSF from one or several suitable bright star(s) (e.g., the
  guide star) for Wiener deconvolution followed by PSF fitting with
  locally extracted PSFs. Both methods lead to satisfying results. The
  method involving Wiener deconvolution is superior. There is a caveat
  to take into account, however. Deconvolution will alter the noise
  statistics of the images. In this work, several tests and
  simulations have shown that this will not lead to any systematic
  bias in the astrometry or photometry, but to an underestimation of
  the uncertainties in the PSF fitting algorithm. This can be taken
  into account by either comparing measurements of several independent
  data sets or by determining a scaling factor for the uncertainties
  from Monte Carlo simulations.

  The method involving Wiener deconvolution is fairly tolerant to the
  PSF used in this primary deconvolution, but it is ideally extracted
  from (a) bright star(s) close to the guide star. In the case of the
  Galactic center and for the data shown here, this approach works
  successfully. Systematic deviations of brightness across the field
  have a standard deviation of $\sim0.02$\,mag. This should be
  compared with the photometric bias that varies systematically across
  the FOV reaching values as high as $\sim0.2$\,mag that occurs when
  using a single PSF for the entire $28"\times28"$ FOV
  (Fig.\,\ref{Fig:dmag}). This corresponds to an improvement in
  accuracy by an order of magnitude. The presented method allows one
  additionally to estimate the diffuse emission due to non-resolved
  sources in a crowded field with an accuracy of $\sim10\%$.

The method was tested on simulated images. The diffuse emission can be
recovered with high accuracy, i.e., deviations of $<10\%$. Of course,
this latter number also depends on the resolution with which the
diffuse emission is estimated. Here we have used a box size of
$30\times30$\,pixels, i.e., about 10 times the FWHM of the PSF or
$0.8"\times0.8"$. Deviations in the estimated diffuse emission can
become larger than 10\% when the PSF used for primary deconvolution is
not determined carefully.

An apparent disadvantage of linear Wiener filter deconvolution is the
typical ringing produced around point sources. However, this is no
problem when a PSF fitting algorithm is applied to the deconvolved
image because the rings around point sources are considered features
of the PSF. Since the Wiener filter deconvolution is a linear
algorithm, information does not become destroyed as in the case of
non-linear methods such as the Lucy-Richardson algorithm, which leads
to deteriorated astrometry and photometry and loss of the information
contained in the diffuse background.

Finally, we emphasize that our experiments have shown
that it is important to take into account the photometric uncertainty
introduced by our limited knowledge of the PSF. While the formal
uncertainty in the point source astrometry and photometry can become
arbitrarily small with the increasing brightness of the source (excluding
saturation), there is a principal limit imposed by the accuracy with
which the PSF can be determined. In the case analyzed here, for
example, the minimum photometric uncertainty due to the PSF may be of the
order $0.01-0.02$\,mag. The minimum PSF uncertainty depends on the
details of the observations, particularly on the availability of
bright, non-saturated PSF reference stars and on the level of
crowding.  We note that the PSF is not constant even in the case of a FOV
smaller than the isoplanatic angle. Therefore, for high precision
astrometry and photometry, the position dependence of the PSF should
always be taken into account when the FOV is not significantly smaller
than the isoplanatic angle.

This work does not intend to provide a readily available standard
solution for PSF fitting with a spatially variable PSF. As has been
mentioned before, convenient program packages already exist for these
cases. The intention of this work was instead to determine ways in
which one can deal with extremely difficult situations, when the
standard methods may not work satisfyingly. My hope is that this work
may also inspire new approaches to the problem in general of variable
PSF fitting.  Further improvement is probably possible by
implementing the true local PSF estimation in {\it StarFinder} and,
possibly, by taking covariances between the pixels into account.
More work on the statistical effects of (Wiener) deconvolution on
astronomical images could also lead to further progress.

The method for obtaining accurate photometry over a large FOV in AO
images with a sparsely sampled PSF, as suggested in this
paper, has been applied to NACO $H$-, $Ks$-, and $Lp$-band imaging
data of the Galactic center. The results (photometry of point sources
and of diffuse emission) will be presented and discussed in an upcoming
paper (Schoedel et al., submitted to A\&A).

\begin{acknowledgements}
RS acknowledges the Ram\'on y Cajal programme of the Spanish
Ministerio de Ciencia e Innovaci\'on. RS thanks the anonymous
reference for his help in improving this work.
\end{acknowledgements}

\bibliography{/Users/rainer_old/Documents/MyPapers/BibGC}

\end{document}